\documentclass[12pt]{article}
 \usepackage{epsfig}
 \def\be{\begin{equation}}
 \def\ee{\end{equation}}
 \def\bea{\begin{eqnarray}}
 \def\eea{\end{eqnarray}}
 \usepackage{graphicx}

 \catcode`\@=11
 \def\lsim{\mathrel{\mathpalette\@versim<}}
 \def\gsim{\mathrel{\mathpalette\@versim>}}
 \def\@versim#1#2{\vcenter{\offinterlineskip
 \ialign{$\m@th#1\hfil##\hfil$\crcr#2\crcr\sim\crcr } }}
 \catcode`\@=12

 \catcode`@=12
\begin{document}
 \thispagestyle{empty}
 \begin{flushright}
 UCRHEP-T577\\
 April 2017\
 \end{flushright}
 \vspace{0.6in}
 \begin{center}
 {\LARGE \bf Pseudo-Majoron as Light Mediator\\ of Singlet Scalar Dark Matter\\}
 \vspace{1.0in}
 {\bf Ernest Ma$^1$ and Markos Maniatis$^2$\\}
 \vspace{0.2in}
 {\sl $^1$ Department of Physics and Astronomy,\\ 
 University of California, Riverside, California 92521, USA\\}
\vspace{0.1in}
{\sl $^2$ Departamento de Ciencias Basicas,\\ Universidad del Biobio, 
Casilla 447, Chillan, Chile\\} 
\end{center}
 \vspace{1.0in}

\begin{abstract}\
In the singlet-triplet majoron model of neutrino mass, lepton number is 
spontaneously broken.  If it is also softly broken, then a naturally 
light pseudoscalar particle $\eta_I$ exists.  It may then act as 
a light mediator for a real singlet scalar $\chi$ with odd dark parity. 
It is itself unstable, but decays dominantly to two neutrinos through 
its triplet scalar component, thereby not disturbing the cosmic microwave 
background (CMB).  It also mixes with the standard-model Higgs boson 
only in one loop, 
thereby not contributing significantly to the elastic scattering of 
$\chi$ off nuclei in dark-matter direct-search experiments. 
\end{abstract}

 \newpage
 \baselineskip 24pt
\noindent \underline{\it Introduction}~:~
In the singlet-triplet majoron model of neutrino mass~\cite{cs91}, 
small Majorana neutrino masses are obtained through a scalar Higgs 
triplet $\Delta = (\Delta^{++},\Delta^+,\Delta^0)$ with lepton number $L=-2$, 
through the Yukawa interactions
\begin{equation}
{\cal L}_Y = -f_{ij}[\nu_i \nu_j \Delta^0 + (\nu_i l_j + l_i \nu_j) \Delta^+/
\sqrt{2} + l_i l_j \Delta^{++}],
\end{equation}
where $f_{ij} = f_{ji}$, resulting in $m^\nu_{ij} = 2 f_{ij} \langle 
\Delta^0 \rangle$.  
In the Higgs potential with the usual $\Phi = (\phi^+,\phi^0)$ doublet of 
the standard model (SM), the trilinear coupling $\Phi^\dagger \Delta \Phi^*$ 
is forbidden by $L$ conservation.  However, if a scalar singlet $\sigma$ 
with $L=2$ is added, then the quadrilinear term
\begin{equation}
\lambda' \sigma \Phi^\dagger \Delta \Phi^* + H.c.
\end{equation}
is allowed.  The spontaneous breaking of $SU(2)_L \times U(1)_Y$ and $L$ 
through the vacuum expectation values $v_{1,2,3}$ as defined by 
\begin{equation}
\sigma = v_1 + {1 \over \sqrt{2}}(\sigma_R + i \sigma_I), ~~~ 
\phi^0 = v_2 + {1 \over \sqrt{2}}(\phi_R + i \phi_I), ~~~ 
\Delta^0 = v_3 + {1 \over \sqrt{2}}(\Delta_R + i \Delta_I), 
\end{equation}
results in four massless Goldstone bosons.  The linear combinations
\begin{equation} 
{v_2 \phi^\pm + \sqrt{2}v_3 \Delta^\pm \over \sqrt{v_2^2 + 2v_3^2}}, ~~~ 
{(v_2 \phi_I + 2 v_3 \Delta_I) \over \sqrt{v_2^2 + 4v_3^2}}
\end{equation}
become the longitudinal component of the $W^\pm$ and $Z$ bosons, and 
the linear combination
\begin{equation}
\eta_I = {[v_1 (v_2^2+4v_3^2)\sigma_I + 2v_3^2 v_2 \phi_I - v_2^2 v_3 
\Delta_I] \over \sqrt{v_1^2 (v_2^2 + 4 v_3^2)^2 + 4 v_2^2 v_3^4 + v_2^4 v_3^2}}
\end{equation}
is the majoron.  If $\sigma$ is absent, this becomes the triplet majoron 
model~\cite{gr81} which is ruled out by $Z$ decay because $\Delta^0$ would 
contribute too much to its invisible width.  Here, assuming that $v_1 >> v_3$, 
this effect can be suppressed.

We now break $L$ also explicitly but softly with the $\sigma^2$ term, 
then the majoron $\eta_I$ becomes massive.  It may be assumed 
naturally light because it is protected by a would-be symmetry.  To 
accommodate dark matter, we add two complex singlet scalars $\chi_{1,2}$ 
which have $L=1$.  The trilinear scalar terms $\chi_i \chi_j \sigma^*$ 
are now allowed.  As a result, $\chi_{1,2}$ have self-interactions through 
the light mediator $\eta_I$, and the enhanced elastic scattering cross 
section~\cite{fkty09} is a possible resolution of the cusp-core anomaly 
in the density profile of dwarf galaxies~\cite{d...09}.  As shown below, 
our model has one very important feature, namely the decay of the majoron 
$\eta_I$ is \underline{dominantly to two neutrinos}.  Its lifetime will 
be very short, and does not disturb the standard big bang nucleosynthesis 
(BBN).  It mixes with the SM Higgs boson only in one loop, and this 
mixing may be arbitrarily small because it is not needed for it 
to decay as in any other model of a light scalar mediator~\cite{m17}. 
This avoids the problem~\cite{kty14} of too large a cross section in 
direct-search experiments.  Further, since $\eta_I$ decays dominantly to two 
neutrinos, it avoids the problem~\cite{gibm09,bksw17} of too much 
disruption to the cosmic microwave background (CMB) if it decays to 
electrons or photons as in all other models of a light scalar mediator.

\noindent \underline{\it Scalar sector}~:~
In our version of the singlet-triplet Majoron model of neutrino mass, 
the Higgs potential is given by
\begin{eqnarray}
V &=& m_1^2 \sigma^* \sigma + m_2^2 \Phi^\dagger \Phi + 
m_3^2 Tr(\Delta^\dagger \Delta) - {1 \over 2} m_4^2 (\sigma^2 + {\sigma^*}^2) 
\nonumber \\ 
&+& {1 \over 2} \lambda_1 (\sigma^* \sigma)^2 + {1 \over 2} \lambda_2 
(\Phi^\dagger \Phi)^2 + {1 \over 2} \lambda_3 [Tr(\Delta^\dagger \Delta)]^2 
+ {1 \over 2} \lambda_4 Tr(\Delta^\dagger \Delta \Delta^\dagger \Delta) 
\nonumber \\ 
&+& \lambda_{12} (\sigma^* \sigma)(\Phi^\dagger \Phi) + \lambda_{13} 
(\sigma^* \sigma)Tr(\Delta^\dagger \Delta) + \lambda_{23} (\Phi^\dagger \Phi) 
Tr(\Delta^\dagger \Delta) + \lambda_{24} \Phi^\dagger \Delta \Delta^\dagger \Phi 
\nonumber \\ 
&-& \lambda' (\sigma \Phi^\dagger \Delta \Phi^* + H.c.)
\end{eqnarray}
where $\sigma$ is a complex neutral singlet and
\begin{equation}
\Phi = \pmatrix{\phi^+ \cr \phi^0}, ~~~ \Delta = \pmatrix{\Delta^{++} & 
\Delta^+/\sqrt{2} \cr \Delta^+/\sqrt{2} & \Delta^0},
\end{equation}
and the $m_4^2$ term has been added to break $L$ softly.  Note that $m_4^2$ 
has been chosen real by rotating the phase of $\sigma$, and $\lambda'$ 
as well by rotating the relative phase of $\Delta$ and $\Phi$. 
Now the minimum of $V$ is determined by
\begin{eqnarray}
0 &=& m_1^2 - m_4^2 + \lambda_1 v_1^2 + \lambda_{12} v_2^2 + \lambda_{13} v_3^2 
- {\lambda' v_2^2 v_3^2 \over v_1}, \\ 
0 &=& m_2^2 + \lambda_2 v_2^2 + \lambda_{12} v_1^2 + (\lambda_{23}+ \lambda_{24}) 
v_3^2 - 2\lambda' v_1 v_3, \\ 
0 &=& m_3^2 + (\lambda_3 + \lambda_4) v_3^2 + \lambda_{13} v_1^2 + (\lambda_{23} 
+ \lambda_{24}) v_2^2 - {\lambda' v_1 v_2^2 \over v_3}. 
\end{eqnarray}
As a result, the $3 \times 3$ mass-squared matrix spanning 
$(\sigma_I,\phi_I,\Delta_I)$ is given by
\begin{equation}
{\cal M}^2_I = \pmatrix{ 2 m_4^2 + \lambda' v_2^2 v_3/v_1 & -2 \lambda' v_2 v_3 & 
\lambda' v_2^2 \cr -2 \lambda' v_2 v_3 & 4 \lambda' v_1 v_3 & -2 \lambda' v_1 v_2 
\cr \lambda' v_2^2 & -2 \lambda' v_1 v_2 & \lambda' v_1 v_2^2/v_3}.
\end{equation}
If $m_4^2=0$, this matrix would have two zero eigenvalues, corresponding to 
the longitudinal component of the $Z$ boson and the majoron of Eq.~(5). 
Removing the former, the reduced $2 \times 2$ matrix spanning $\sigma_I$ 
and $(v_2 \Delta_I - 2v_3 \phi_I)/\sqrt{v_2^2+4v_3^2}$ becomes
\begin{equation}
{\cal M}^2_I = \pmatrix{ 2 m_4^2 + \lambda' v_2^2v_3/v_1 &  \lambda' v_2 
\sqrt{v_2^2 + 4 v_3^2} \cr \lambda' v_2 \sqrt{v_2^2 + 4 v_3^2} & 
\lambda' v_1 (v_2^2 + 4 v_3^2)/v_3}.
\end{equation}
We know that $v_3$ has to be small compared to $v_2$ from precision 
electroweak data.  We know also that $v_3$ has to be small compared to 
$v_1$ because the triplet majoron is ruled out from the measurement of 
the $Z$ invisible width.  Hence the mixing between the two states 
is small, i.e. $v_3/v_1$, with the two physical states having the 
squares of their masses equal to $2m_4^2$ for the light pseudo-majoron 
$\eta_I$ and $\lambda' v_1 v_2^2/v_3$ for the other scalar which is 
heavy.

In the $(\sigma_R,\phi_R,\Delta_R)$ sector for $v_3 << v_{1,2}$, $\Delta_R$ 
has the same mass-squared as $\Delta_I$, i.e. $\lambda' v_1 v_2^2/v_3$, 
whereas $\sigma_R$ and $\phi_R$ mix according to
\begin{equation}
{\cal M}^2_R = \pmatrix{2 \lambda_1 v_1^2 & 2 \lambda_{12} v_1 v_2 \cr 
2 \lambda_{12} v_1 v_2 & 2 \lambda_2 v_2^2}.
\end{equation}
This means that the observed 125 GeV scalar at the Large Hadron Collider 
(LHC) may have a small singlet component which does not couple to quarks or 
leptons.

\noindent \underline{\it Dark sector}~:~
Consider now the addition of two complex neutral singlet scalars $\chi_{1,2}$ 
with $L=1$.  We add to $V$ the following scalar potential:
\begin{eqnarray}
V' &=& m_5^2 \chi_1^* \chi_1 + m_6^2 \chi_2^* \chi_2 + {1 \over 2} \lambda_5 
(\chi_1^* \chi_1)^2 + {1 \over 2} \lambda_6 (\chi_2^* \chi_2)^2 + 
\lambda_7 (\chi_1^* \chi_1)(\chi_2^* \chi_2) \nonumber \\ 
&+& \lambda_{15} (\sigma^* \sigma) (\chi_1^* \chi_1) + \lambda_{16} 
(\sigma^* \sigma) (\chi_2^* \chi_2) + [\lambda_{17} (\sigma^* \sigma)
(\chi_1^* \chi_2) + H.c.] \nonumber \\  
&+& \lambda_{25} (\Phi^\dagger \Phi) (\chi_1^* \chi_1) + \lambda_{26} 
(\Phi^\dagger \Phi) (\chi_2^* \chi_2) + [\lambda_{27} (\Phi^\dagger \Phi)
(\chi_1^* \chi_2) + H.c.] \nonumber \\  
&+& \lambda_{35} Tr(\Delta^\dagger \Delta) (\chi_1^* \chi_1) + \lambda_{36} 
Tr(\Delta^\dagger \Delta) (\chi_2^* \chi_2) + [\lambda_{37} Tr(\Delta^\dagger \Delta)
(\chi_1^* \chi_2) + H.c.] \nonumber \\   
&+& \left[{1 \over 2} \mu_{11} \chi_1^2 \sigma^* + {1 \over 2} \mu_{22} 
\chi_2^2 \sigma^* + \mu_{12} \chi_1 \chi_2 \sigma^* + H.c.\right] 
\end{eqnarray}
The sum $V + V'$ is invariant under $U(1)_L$ in all its dimesnion-four 
and dimension-three terms.  The only term which breaks $U(1)_L$ explicitly 
is the dimesnion-two $m_4^2$, so that the symmetry of $V + V'$ becomes 
$Z_4$, under which the charges of $\Phi, \sigma, \Delta, \chi_{1,2}$ are 
$1, -1, -1, i, i$ respectively.  Once the spontaneous breaking of $V + V'$ 
occurs with $v_{1,2,3}$, then the residual symmetry becomes $Z_2$, under 
which $\Phi, \sigma, \Delta$ are even and  $\chi_{1,2}$ odd.  Hence the 
lightest $\chi$ is a dark-matter candidate.  Note that if only one copy 
of $\chi$ is used, $V + V'$ would have only real parameters, and there 
would not be a trilinear coupling linking $\chi$ to the light would-be 
pseudoscalar majoron.  Hence the dark matter in this case would not 
have any enhanced self-interactions.  Note also that this is another 
example~\cite{m15} of the derivation of dark parity from lepton parity, 
i.e. $(-1)^{L+2j}$ from $(-1)^{L}$. 

In Eq.~(14), we can rotate the phases of $\chi_{1,2}$ to make 
$\mu_{11}, \mu_{22}$ real, then $\mu_{12}$ remains complex, as well as 
$\lambda_{17}, \lambda_{27}, \lambda_{37}$.  The mass-squared matrix 
spanning $\chi_{1R}, \chi_{2R}, \chi_{1I}, \chi_{2I}$ is then given by
\begin{equation}
{\cal M}^2_\chi = \pmatrix{ m_5^2 + \mu_{11} v_1 + \Lambda_5 & \mu^R_{12} v_1 + 
\Lambda^R_7 & 0 & -\mu^I_{12} v_1 - \Lambda^I_7 \cr \mu^R_{12} v_1 + \Lambda^R_7 
& m_6^2 + \mu_{22} v_1 + \Lambda_6 & -\mu^I_{12} v_1 + \Lambda^I_7 & 0 
\cr 0 & -\mu^I_{12} v_1 + \Lambda^I_7 & m_5^2 - \mu_{11} v_1 + \Lambda_5 & 
-\mu^R_{12} v_1 + \Lambda^R_7 \cr -\mu^I_{12} v_1 - \Lambda^I_7 & 0 & 
-\mu^R_{12} v_1 + \Lambda^R_7 & m_6^2 - \mu_{22} v_1 + \Lambda_6},
\end{equation}
where $\Lambda_5 = \sum \lambda_{i5} v_i^2$, $\Lambda_6 = \sum \lambda_{i6} 
v_i^2$, $\Lambda^R_7 = \sum \lambda^R_{i7} v_i^2$, $\Lambda^I_7 = \sum 
\lambda^I_{i7} v_i^2$.  Now $\sigma_I$ couples to the matrix
\begin{equation}
{\cal Y}_I = {1 \over 2 \sqrt{2}}\pmatrix{ 0 & \mu^I_{12} & \mu_{11} 
& \mu^R_{12} \cr \mu^I_{12} & 0 & \mu^R_{12} & \mu_{22} \cr \mu_{11} 
& \mu^R_{12} & 0 & -\mu^I_{12} \cr \mu^R_{12} & \mu_{22} & -\mu^I_{12} & 0}.
\end{equation}
This shows that if $\mu^I_{12}$ and $\Lambda^I_7$ were zero, then 
$\sigma_I$ couples only to two different physical $\chi$ states, so that 
there is no tree-level self-interaction of dark matter through $\sigma_I$.  
Let the lightest eigenstate of Eq.~(15) 
be $\chi_0 = a \chi_{1R} + b \chi_{2R} + c \chi_{1I} + d \chi_{2I}$, then 
the $\sigma_I \chi_0 \chi_0$ coupling is
\begin{equation}
\mu_0^I = {1 \over \sqrt{2}}[ac \mu_{11} + bd \mu_{22} + (ad+bc) \mu^R_{12} 
+ (ab - cd) \mu^I_{12}].
\end{equation}
Similarly, $\sigma_R$ couples to the matrix
\begin{equation}
{\cal Y}_R = {1 \over 2 \sqrt{2}}\pmatrix{ 2 \lambda_{15} v_1 + \mu_{11} 
& 2 \lambda^R_{17} v_1 + \mu^R_{12} & 0 
& -2 \lambda^I_{17} v_1 -\mu^I_{12} \cr 2 \lambda^R_{17} v_1 + \mu^R_{12} 
& 2 \lambda_{16} v_1 + \mu_{22} & 2 \lambda^I_{17} v_1 -\mu^I_{12} 
& 0 \cr 0 & 2 \lambda^I_{17} v_1 -\mu^I_{12} & 2 \lambda_{15} v_1 -\mu_{11} 
& 2 \lambda^R_{17} v_1 -\mu^R_{12} 
\cr -2 \lambda^I_{17} v_1 -\mu^I_{12} & 0 & 2 \lambda^R_{17} v_1 -\mu^R_{12} 
& 2 \lambda_{16} v_1 -\mu_{22}}.
\end{equation}
Hence the $\sigma_R \chi_0 \chi_0$ coupling is
\begin{equation}
\mu^R_0 = {1 \over \sqrt{2}} \left[ {1 \over 2}(a^2-c^2)\mu_{11} + 
{1 \over 2}(b^2-d^2) \mu_{22} + (ab-cd)\mu^R_{12} - (ad+bc)\mu^I_{12} \right] 
+ 2 \sqrt{2} \lambda_{\sigma \chi} v_1,
\end{equation}
where $\lambda_{\sigma \chi}$ is the $(\sigma_R^2 + \sigma_I^2) \chi_0^2$ 
coupling given by
\begin{equation}
\lambda_{\sigma \chi} = {1 \over 4}[(a^2+c^2)\lambda_{15} + 
(b^2+d^2)\lambda_{16} + 2(ab+cd)\lambda^R_{17} - 2(ad-bc)\lambda^I_{17}]
\end{equation}

\noindent \underline{\it Dark matter interactions}~:~
In our model, $\chi_0$ is dark matter and the pseudo-majoron $\eta_I$ (mostly 
$\sigma_I$) is its light mediator.  Since $m^2_{\eta_I} << m^2_{\chi_0}$, 
the $s$-channel contribution is suppressed, and the elastic scattering of 
$\chi_0$ at rest, through the exchange of $\eta_I$ in the $t$ and $u$ 
channels, is enhanced and given by
\begin{equation}
\sigma(\chi_0 \chi_0 \to \chi_0 \chi_0) = {(\mu_0^I)^4 \over 16 \pi 
m^4_{\eta_I} m^2_{\chi_0}}.
\end{equation}
For the benchmark value of $\sigma/m_{\chi_0} \sim 1~cm^2/g$ for 
self-interacting dark matter, it may be satisfied with
\begin{equation}
m_{\chi_0} = 100 ~{\rm GeV}, ~~~ m_{\eta_I} = 10~{\rm MeV}, ~~~ 
\mu_0^I = 7~{\rm GeV}.
\end{equation}

Consider now the annihilation of $\chi_0 \chi_0 \to \eta_I \eta_I$. 
If only the $\mu_I$ interaction is used, then this cross section is 
much smaller than the canonical value $\sigma_0 \times v_{rel} = 
3 \times 10^{-26}~cm^3/s$ for the correct dark matter relic abundance of 
the Universe.  However, $\chi_0$ annihilation to $\eta_I$ may also proceed 
through $\mu_0^R$ and $\lambda_{\sigma \chi}$.  Assuming the latter to be 
dominant, we find
\begin{equation}
\sigma(\chi_0 \chi_0 \to \eta_I \eta_I) \times v_{rel} = 
{\lambda^2_{\sigma \chi} \over 32 \pi m^2_{\chi_0}},
\end{equation}
which works for $\lambda_{\sigma \chi} = 0.05$.
The dark matter scalar $\chi_0$ also couples to the SM Higgs boson $h$ 
(mostly $\phi_R$) through the $\lambda_{25}$, $\lambda_{26}$, and 
$\lambda_{27}$ terms of Eq.~(14).  Hence direct $\chi_0$ annihilation 
to SM particles is also possible.  They have been neglected here for 
simplicity.  If they are nonnegligible, they could be important for 
the indirect detection of $\chi_0$ in space.  As for $\eta_I$, although 
it does not mix with $h$ at tree 
level, there is an allowed $\eta_I \eta_I h$ coupling which will keep 
it in thermal equilibrium with the SM particles until it decays away.

\noindent \underline{\it Decay of the pseudo-majoron}~:~
As shown in Ref.~\cite{m17}, $\eta_I$ mixes with $h$ through $\chi_{1,2}$ 
in one loop.  This phenomenon of radiative Higgs mixing has only been 
discovered recently~\cite{m16}.  If this were the dominant decay mode 
of $\eta_I$, then its decay product, i.e. $e^- e^+$, would disturb the 
CMB, and because of the large Sommerfeld enhancement~\cite{s31} for 
late-time decays, this effect would rule out~\cite{bksw17} any 
self-interacting dark matter with $s$-wave annihilation which is strong 
enough to address the small-scale problems of structure formation. 
There is also an important constraint~\cite{kty14} from direct-search 
experiments. 
For $m_{\chi_0} \sim 100$ GeV, the nonobservation of dark matter so far 
places a bound on the $\eta_I - h$ mixing, which makes the $\eta_I$ 
lifetime too long to accommodate the success of the standard BBN.

Both problems are solved here because $\eta_I$ decays to two neutrinos 
at tree level through its $\Delta_I$ component.  Using Eqs.~(1) and (5), 
the $\eta_I \nu_i \nu_j$ coupling is $-f_{ij} v_3/v_1$.  The decay 
rate of $\eta_I$ to $\nu_i \nu_j$ and $\bar{\nu}_i \bar{\nu}_j$ is then
\begin{equation}
\Gamma = {m_{\eta_I} \sum_{ij} |m^\mu_{ij}|^2 \over 16 \pi v_1^2}.
\end{equation}
Setting $m_{\eta_I} = 10$ MeV, $\sum_{ij} |m^\mu_{ij}|^2 = 1$ eV$^2$, and 
$v_1 = 10$ GeV, we find $\Gamma^{-1} = 0.33~s$ which is less than the 
benchmark of $1~s$ for the $\eta_I$ lifetime not to be a problem for 
the standard BBN.  The $\eta_I - h$ mixing may now be chosen to be 
negligible, so the direct-search bound is not applicable, and since 
$\eta_I$ decays dominantly to neutrinos, the strong constraints of 
the CMB are also avoided.

\noindent \underline{\it Phenomenological consequences}~:~
As shown in a previous section, all the components of the scalar triplet 
$(\Delta^{++},\Delta^+,\Delta^0)$ are heavy with mass squared 
roughly $\lambda' v_1 v_2^2/v_3$.  Three scalar particles remain: the SM 
Higgs $h$ which is mostly $\phi_R$, the pseudo-majoron $\eta_I$ which is 
mostly $\sigma_I$, and the orthogonal scalar to $h$ which is mostly 
$\sigma_R$.  For $v_1 \sim 10$ GeV, $m_{\sigma_R} \sim 1$ GeV is expected, 
in which case $h \to \sigma_R \sigma_R$ is possible through $\lambda_{12}$ 
of Eq.~(6).  The subsequent decay of $\sigma_R$ through its mixing with $h$ 
to SM particles may be searched for~\cite{cmy14} at the LHC.

The dark sector has two complex scalars $\chi_{1,2}$ where all four components 
mix as shown in Eq.~(15).  The lightest mass eigenstate $\chi_0$ is dark 
matter.  It couples to the pseudo-majoron $\eta_I$ through the trilinear 
term $\mu^0_I \eta_I \chi_0^2$ and $\lambda_{\sigma \chi} \eta_I^2 \chi_0^2$.
For $m_{\eta_I} = 10$ MeV, $m_{\chi_0} = 100$ GeV, and $\mu^I_0 = 7$ GeV, 
the elastic scattering cross section of $\chi_0$ is large enough to 
explain the cusp-core discrepancy of the density profile of dwarf 
galaxies.  For $\lambda_{\sigma \chi} = 0.05$, the annihilation of $\chi_0$ 
to $\eta_I$ is also just right for it to account for the observed relic 
abundance of dark matter.  The decay of $\eta_I$ is dominantly to two 
neutrinos.  It does not disturb the standard BBN or the CMB.

The mixing of $\eta_I$ with $h$ is small.  For a given $m_{\eta_I}$, it 
is constrained by direct-search experiments.  However, since its value 
is unknown, it may still be large enough for $\chi_0$ to be detected 
in underground experiments through $\eta_I$ exchange in the future.
The occasional annihilation of $\chi_0$ in space produces $\eta_I$, 
but since the latter decays dominantly to neutrinos, it will be 
difficult to observe in satellite or ground-based experiments.

\noindent \underline{\it Concluding remarks}~:~
In the singlet-triplet majoron model of neutrino mass, a light 
pseudo-majoron $\eta_I$ is natural and can be chosen for the light mediator 
of self-interacting dark matter $\chi_0$ based on the conservation of lepton 
parity extended to dark parity.  The important property of $\eta_I$ is 
that \underline{it decays dominantly to neutrinos}, thus avoiding strong 
constraints 
from the CMB, as well as the potential conflict between direct-search bounds 
and the standard BBN.  We also predict a singlet scalar $\sigma_R$ of 
about 1 GeV, which mixes with the standard Higgs boson $h$.  From 
$h \to \sigma_R \sigma_R$ decay, it may be discovered at the LHC. 
The $\eta_I - h$ mixing may also allow underground experiments to 
discover $\chi_0$, but the annihilation of $\chi_0$ in space to $\eta_I$ 
would not be easy to detect.  However, the $\chi_0 \chi_0$ cross section 
to SM particles may be significant through the Higgs portal and could 
provide a means for its discovery. 

\noindent \underline{\it Acknowledgements}~:~
This work is supported in part by the U.~S.~Department of Energy under 
Grant No.~DE-SC0008541 and by Fondecyt (Chile) Grant No. 1140568.

\bibliographystyle{unsrt}

\end{document}